# Primary oscillatory instability in low-aspect-ratio rotating disk – cylinder system (rotor – stator cavity)

A. Yu. Gelfgat

*School of Mechanical Engineering, Faculty of Engineering, Tel Aviv University, Tel-Aviv, 69978, Israel*

**Abstract**

Three-dimensional instability of axisymmetric flow in a rotating disk – cylinder configuration is studied numerically for the case of low cylinders with the height/radius aspect ratio varying between 1 and 0.1. A complete stability diagram for the transition from steady axisymmetric to oscillatory three-dimensional flow regime is reported. A good agreement with experimental results is obtained. It is shown that critical azimuthal wavenumber grows with the decrease of the aspect ratio, reaching the value of 19 at the aspect ratio 0.1. It is argued that the observed instability cannot be described as resulting from a Bödewadt flow or from a boundary layer only. Other reasons that can destabilize the flow are discussed.

## 1. Introduction

Flow in a cylinder covered by a rotating disk has been one of most popular swirling flow models during several last decades. Early studies were mainly devoted to phenomenon of vortex breakdown that the flow exhibits under certain conditions[1-3]. Later, attention was focused on stability of base axisymmetric flow[4,5], non-linear development of instabilities[9], stability control[10-13], and effects of some mechanical, thermal or electromagnetic flow complications (not cited here). The flow is governed by two parameters: Reynolds number $Re$ based on the disk radius and angular velocity and height/radius aspect ratio $\gamma$. The early computational studies of flow stability[4,5] considered moderate aspect ratios $1 \leq \gamma \leq 3.5$ that corresponded to classical experiments of Escudier[1]. These results were verified numerically in several independent studies and later were validated experimentally by Sørensen et al[14], so that their correctness seems to be proved. A similar cross-validation of numerical and experimental results was repeated later by Sørensen et al[15] for higher cylinders with the aspect ratio reaching $\gamma = 6$.

Along with a large number of studies considering aspect ratios larger than unity, a group of authors[16-27] examined the rotating disk – cylinder flow at small aspect ratios, mainly at $\gamma = 0.25$ and 0.1. These small aspect ratio configurations sometimes are considered as an extension of the classical von Karman rotating disk in stationary fluid, and Bödewadt stationary disk in rotating fluid flows.[23,26] Several authors emphasize connection between the model and flows between rotating compressors and turbine disks,[19,26] calling the model "rotor-stator cavity" or "rotor-stator system". Main results of these studies are reviewed by Launder et al[28]. In this study we are most interested in results obtained for instability of the base axisymmetric flow in rotor-stator cavities, a brief examination of which shows that the numerical results[19,20,22,26] do not agree between themselves, as well as do not match corresponding experimental findings[21,27]. This disagreement motivated the present study whose main goal is extension of our previous stability results to small aspect ratio cylinders up to $\gamma = 0.1$.

In the following we examine three-dimensional instability of the axisymmetric base flow in the rotating disk – cylinder system for aspect ratios decreasing from 1 to 0.1. The linear stability analysis separates for different azimuthal Fourier modes, so that we find the most critical circumferential periodicity (Fourier mode). In particular, we show that at $\gamma = 0.1$ the critical circumferential periodicity is $2\pi/19$, and argue that this fact only can lead to certain numerical difficulties when a numerical grid is used. Present numerical results are successfully compared with the experiments[21,27]. To gain some more insight in reasons causing flow destabilization, we present patterns of flows and most unstable perturbations and perform some additional estimates and numerical experiments. This allows us to discuss reasons that make the flow unstable. We argue also that existence of multiple supercritical oscillatory three-dimensional flow states should be expected already at small supercriticalities.

## 2. Formulation of the problem and numerical method

We consider a flow of Newtonian incompressible liquid with density ρ and kinematic viscosity ν, in a vertical cylindrical container of radius $R$ and height $H$, whose upper cover rotates around its axis with the angular velocity Ω. The flow is governed by the momentum and continuity equations that are rendered dimensionless using $R$, $R^2/\nu$, $\Omega R$, and $\rho(\Omega R)^2$ as scales of length, time, velocity and pressure, respectively. The dimensionless equations and no-slip boundary conditions are defined in a cylinder $0 \leq z \leq \gamma$, $0 \leq r \leq 1$, and reads

$$\frac{\partial \boldsymbol{v}}{\partial t} + (\boldsymbol{v} \cdot \nabla)\boldsymbol{v} = -\nabla p + \frac{1}{Re}\Delta \boldsymbol{v}, \quad \nabla \cdot \boldsymbol{v} = 0, \qquad (1,2)$$

$$\boldsymbol{v}(r, z = 0) = \boldsymbol{v}(r = 1, z) = \boldsymbol{0}, \qquad (3)$$

$$v_r(r, z = \gamma) = v_z(r, z = \gamma) = 0, \quad v_\theta(r, z = \gamma) = r. \qquad (4)$$

Here $p$ is the pressure, $\boldsymbol{v} = \{v_r, v_\theta, v_z\}$ is the velocity, $\gamma = H/R$ is the aspect ratio of the cylinder, and $Re = \Omega R^2/\nu$ is the Reynolds number. The problem is formulated in cylindrical coordinates, so that the velocity and pressure fields are $2\pi$ – periodic with respect to the polar angle $\theta$. In the following we assume that the base flow state $\boldsymbol{V} = \{V_r(r,z), V_\theta(r,z), V_z(r,z)\}$ is steady and axisymmetric, and consider its stability with respect to infinitesimally small three-dimensional disturbances defined as $\boldsymbol{u} = \{u_r(r,z), u_\theta(r,z), u_z(r,z)\}exp(\lambda t + ik\theta)$. After the base flow is calculated, the linear stability problem reduces to an eigenvalue problem for the time increment $\lambda$ that separates for each integer azimuthal wavenumber $k$. For each aspect ratio $\gamma$ we calculate marginal values of the Reynolds number $Re_m(k)$ at which the leading eigenvalue (i.e., eigenvalue with the largest real part) crosses the imaginary axis. The critical Reynolds number $Re_{cr}$ is defined as the minimal $Re_m(k)$ over all values of $k$. More details on the problem formulation can be found in Refs. 2-5.

The problem is solved using finite volume discretization in space and numerical approach described in Ref. 29. The numerical method was already verified for arbitrary values of the aspect ratio[29] and validated against experimental results for $3.3 \leq \gamma \leq 5.5$.[15] Therefore, we needed to examine the convergence mainly for the smallest value of the aspect ratio, $\gamma = 0.1$, considered below. According to our previous conclusions[29], a calculation of critical Reynolds number needs at least 100 grid points in the shortest spatial direction, while results can be further improved using the Richardson extrapolation. Therefore, for results reported below we performed two independent calculations on two grids with the number of axial points $N_z = 190$ and 200, and number of radial points $N_r = N_z/\gamma$, followed by the Richardson extrapolation of the two results. Results of calculations on the two grids differ only in the fifth decimal place. After Richardson extrapolation is done four first decimal places are considered as converged.

## 3. Results

Main stability results are shown in Figures 1 and 2 for $0.2 \leq \gamma \leq 1$ and $0.1 \leq \gamma \leq 0.2$, respectively. Marginal Reynolds numbers $Re_m$ calculated for different azimuthal wavenumbers $k$ are shown in frames (a). As mentioned above, at $Re = Re_m$ the largest real part of all the eigenvalues $\lambda(k)$ is zero, and the corresponding eigenvalue and eigenmode are called leading. The imaginary part of leading eigenvalue $\omega_m$ can be interpreted as a marginal value of the oscillations circular frequency, and is called marginal frequency. The marginal frequencies are shown in frames (b) of Figs. 1 and 2. The disturbances are proportional to $exp(ik\theta + i\omega_m t)$, so that they appear as waves travelling in the azimuthal direction with the phase angular velocity $\omega_m/2\pi k$. The disk is assumed to rotate with a positive angular frequency $\Omega$ in the (positive) counter-clockwise direction. Therefore, negative values of $\omega_m$ correspond to a wave travelling in the (negative) clockwise direction, opposite to the disk rotation. The parameter $a = R/2H$ used in Refs. 19,20,25,26 instead of the aspect ratio $\gamma$ is shown as the upper horizontal axis of frames (a).

Looking at the critical values of the Reynolds numbers $Re_{cr}$ represented by the lower envelope of all curves in the frames (a), we see that at $\gamma = 1$ the instability sets in with the azimuthal number $k_{cr} = 2$ (Fig. 1a), as predicted in Ref. 5 by another numerical technique, and confirmed experimentally in Ref. 10. With the decrease of aspect ratio the critical azimuthal wavenumber $k_{cr}$ grows, becoming $k_{cr} = 3$ at $\gamma \approx 0.85$, $k_{cr} = 4$ at $\gamma \approx 0.5$, and $k_{cr} = 5$ at $\gamma \approx 0.4$ (Fig. 1a). With further decrease of the aspect ratio the critical azimuthal wavenumber grows even more rapidly, reaching $k_{cr} = 10$ at $\gamma \approx 0.2$, and $k_{cr} = 19$ at $\gamma \approx 0.1$ (Fig. 2a). At these large values of $k$ the marginal values of $Re_m(\gamma)$ are very close. The differences between marginal Reynolds numbers become so small that they can be distinguished by comparing the numbers, but not by comparing the curves. To illustrate this, marginal Reynolds numbers and frequencies at $\gamma = 0.1$ and $15 \leq k \leq 25$ are reported in Table 1. We observe that the critical Reynolds number at $k = 19$ is less than one per cent smaller than $Re_m$ at several neighbour values of $k$. This means that selection of the azimuthal periodicity in supercritical flow regimes is dependent on the initial conditions, so that multiplicity of asymptotic oscillatory state at $Re > Re_{cr}$ can be expected. The curves of $\omega_m(\gamma)$ at different azimuthal wavenumbers are rather well separated (frames (b) in Figs. 1 and 2), so that frequency value can serve as indication of a certain azimuthal periodicity, as it was done in Refs. 10 and 15.

Another important conclusion that follows from the large azimuthal wavenumber instability observed at small aspect ratios relates to requirements that any computational modelling of supercritical oscillatory flows should meet. To resolve the instability at $k_{cr} = 19$, the numerical method should account for at least 20 first Fourier harmonics ($0 \leq k \leq 19$). To account for the main non-linear effects the number of harmonics should be doubled. In case of numerical method using a grid in the circumferential direction, the number of grid nodes must be sufficient to resolve structures located within a sector defined by the angle of $2\pi/19$. Assuming a coarse resolution of 10 points per structure, we arrive to 190 azimuthal grid points. None of previous calculations, done for $\gamma = 0.1$, met these accuracy requirements (see next Section).

Flow and leading perturbation patterns are illustrated in Figs. 3-5. The flow is shown by streamlines of the meridional flow (upper left frames) and isolines of the azimuthal velocity $V_\theta$ (upper right frames). Amplitude of the leading (or most unstable) perturbation $\boldsymbol{u}(r,z) = \{u_r(r,z), u_\theta(r,z), u_z(r,z)\}$ is a complex vector function, and we show real and imaginary part together with the absolute value of all three components. Since the leading perturbation is an eigenvector of the linear stability problem, it is defined within multiplication by a constant. This means that values of maximal moduli reported in the figures can be used only by comparison of maxima of different perturbation components between themselves. Thus, we observe that perturbation of the azimuthal velocity is always larger than that of the two other velocity components. The absolute value of the perturbations represent distribution of the amplitude of flow oscillations in the meridional plane, while patterns of its real and imaginary parts correspond to two momentarily snapshots taken at time interval of a half oscillations period. These two patterns are defined within multiplication by a real constant and a choice of their complex phase. Note that small structures of real and imaginary parts of perturbations, appearing along the cylinder bottom in Figs. 4 and 5 require better spatial resolution than the base flow itself does. This is the main reason for applying fine grids, up to 2000×200 nodes, in the current study.

Patterns of the most unstable perturbations (Figs. 3-5) allow one to make several conclusions and some reasonable speculations about flow physics involved in the instability onset. First, we observe that the perturbations are most intensive near the stationary bottom of the cylinder. These patterns show that the instability observed is not related to instabilities of von Karman flow near rotating disk. Localization of the most unstable perturbation near the stationary lower disk (Figs. 4 and 5) recalls instability of the Bödewadt flow, as well as instability of boundary layer adjacent to the stationary disk. Instabilities of the two latter models were studied in Serre et al,[26] who found that in both cases the instabilities set in as travelling waves rotating in the direction opposite to the upper disk rotation. This qualitatively agrees with the present negative values of $\omega_m$. Furthermore, the instabilities were predicted to set in at local Reynolds number $Re_\delta < 50$, where $Re_\delta = r^*\sqrt{\Omega/\nu}$, and $r^*$ is the local distance to the axis. Using a rough estimation $r^* = 0.5R$ and $Re_{cr} \approx 2 \cdot 10^4$ (Fig. 2), we estimate $Re_\delta \approx 0.5\sqrt{Re} \approx 70$, which also fits the theoretical prediction. We could not find, however, how the theoretical results of Ref. 26 can predict a large critical azimuthal wavenumber. Also, the above arguments do not relate to the perturbation patterns shown in Fig. 3 for $\gamma = 0.6$. In this case we observe also large perturbation amplitude near the cylinder wall. A closer look on the disturbances moduli shown in Figs. 4 and 5 also reveals a possible effect of lateral wall on the flow instability.

Looking for another possible explanation of instability we notice again that the most intensive disturbances are located near the stationary bottom, where the main meridional circulation is directed from the cylinder wall towards the axis. This reminds a well-known instability mechanism where azimuthal velocity perturbation is advected towards the rotation axis and increases owing to conservation of the rotational momentum. Such instability takes place in inviscid rotating flows when the Rayleigh criterion $\frac{\partial}{\partial r}(rV_\theta)^2$ becomes negative. Examination of the Rayleigh criterion shows that it is positive along most of the stationary bottom. It is strongly negative along the stationary cylinder wall, approaching which the azimuthal velocity steeply decreases. The largest negative magnitudes are located near the upper corner where the azimuthal velocity is discontinuous, however, we do not observe any strong disturbances there. Slightly negative values of the Rayleigh criterion persist in the lower corner of the cylinder, where we do observe non-negligible perturbation amplitudes. The latters are noticeably strong at $\gamma = 0.6$ and 0.1 (Figs. 3 and 5) and rather weak at $\gamma = 0.25$ (Fig. 4). Since viscosity effects are necessarily strong near the corners, and changes of the meridional velocities are steeper than that of the rotational one, the Rayleigh criterion is hardly applicable there.

We tried also to use approach of Ref. 30, where different terms of linear stability problem are cancelled one by one, while observing cancellation of which of them leads to flow stabilization. Here "flow stabilization" means a change of sign of the leading eigenvalue from positive to negative. This exercise shows that cancellation of the terms $V_r\frac{\partial u_r}{\partial r}$, $V_r\frac{\partial u_\theta}{\partial r}$, and $V_r\frac{\partial u_z}{\partial r}$ does not yield any stabilization, which causes additional doubts about instability of the boundary layer, in which the radial base velocity is dominant. At the same time it was observed that the strongest stabilization is achieved when the term $V_z\frac{\partial u_\theta}{\partial z}$ is excluded. Note that $V_z$ is large near the sidewall, but is very small near the bottom stationary disk. Cancellation of several other terms (including those corresponding to perturbations of the centrifugal and Coriolis forces) provided weaker stabilization. Taking into account the perturbation patterns (Figs. 3-5) we can speculate that a disturbance of the azimuthal velocity is advected along the vertical wall towards the bottom, where the meridional flow changes its direction and drives the perturbations towards the axis. Along this motion the perturbation of the azimuthal velocity necessarily grows owing to conservation of the rotational momentum. This growth affects the centrifugal and Coriolis forces. The centrifugal force acts against the flow directed towards the axis, which can be a source of flow oscillations. The Coriolis force drives the perturbations in azimuthal direction, which can lead to appearance of the azimuthal travelling wave.

## 4. Comparison with experiments and previous calculations

Transition from steady to oscillatory flow regime was studied experimentally in Ref. 21 for $\gamma = 0.1$ and in Ref. 27 for $\gamma = 0.114$. Since the two aspect ratios are close, so are the critical Reynolds numbers that are measured to be 25,000 for $\gamma = 0.1$, and 20,500 for $\gamma = 0.114$. The frequency of oscillating supercritical regime was reported only in Ref. 27 as $\omega = 2.1$ for $\gamma = 0.114$. The results of present calculations yield at $\gamma = 0.1$: $Re_{cr} = 22,350$, $\omega_{cr} = 2.426$, $k_{cr} = 19$; and for $\gamma = 0.114$: $Re_{cr} = 19,580$, $\omega_{cr} = 2.108$, $k_{cr} = 17$. Thus, our numerical results are in a good agreement with experimental measurements. The experimental critical values of the Reynolds number are slightly above the numerical ones, as one would expect for a supercritical bifurcation.

Comparison with the previously published numerical results appears to be not so successful. Thus, results reported for $\gamma = 0.25$ in Refs. 19,20 is $Re_{cr} = 16,000$, $\omega_{cr} \approx 0.9$. The critical azimuthal wavenumber was not reported. The present result is $Re_{cr} = 7,960$, $\omega_{cr} = 0.395$, $k_{cr} = 7$. The cited calculations used only 48 grid nodes in the azimuthal direction, which left only 6-7 grid nodes per azimuthal period at $k = 7$. This is clearly insufficient and can be the main reason for the disagreement. Other unsuccessful comparison is made for $\gamma = 0.1$. The previously published results are $Re_{cr} = 35,000$, $\omega_{cr} \approx 1.0$ (Ref. 20), and $Re_{cr} = 12,000$, $\omega_{cr} \approx 1.5$ (Ref. 26). These critical Reynolds numbers are above and below the one reported here, $Re_{cr} = 22,350$ (see above paragraph). Both calculations used 48 grid points in the azimuthal direction, which does not seem sufficient already for $\gamma = 0.25$. Obviously with a such an insufficient number of points it was impossible to resolve the instability at $k_{cr} = 19$.

## 5. Conclusions

Study of three-dimensional instability of flow in rotating disk – cylinder system is extended to low aspect ratio containers, up to the height/radius ratio 0.1. The main stability results are reported as dependencies of the marginal Reynolds numbers and marginal frequencies on the aspect ratio and the azimuthal wavenumber $k$. It is found that the critical azimuthal wavenumber grows with the decrease of the aspect ratio $\gamma$, so that $k_{cr} = 10$ at $\gamma \approx 0.2$, and $k_{cr} = 19$ at $\gamma \approx 0.1$. At small aspect ratios we observe close marginal values of the Reynolds number corresponding to $k$ close to $k_{cr}$, which can lead to multiplicity of stable oscillatory supercritical states. The azimuthal modes can be distinguished, in particular, by their oscillation frequencies, whose marginal values remain noticeably different at different $k$.

A good agreement with the experimental measurements of instability at $\gamma = 0.1$ and $0.114$ are reported, which, together with the results of Refs. 10 and 15, completes the experimental validation of our present and previous stability results. At the same time, present results differ noticeably from the previously published numerical ones. We argue that the disagreement follows mainly from insufficient numerical resolution in the azimuthal direction, so that the previous studies could not resolve instabilities at $k \geq 7$.

Patterns of the most unstable perturbations show that at large aspect ratios the instability sets in along the stationary cylinder bottom. After several estimations and numerical experiments we argued that the observed instability cannot be described solely as instability of the Bödewadt flow, or instability of the boundary layer attached to the stationary disk. We argue that the instability sets in owing to perturbations of the azimuthal velocity that are advected along the sidewall towards the bottom, where they increase due to conservation of the angular momentum and then affect the radial and circumferential fluid motion.

**Figure captions**

Fig. 1. Marginal Reynolds numbers (a) and marginal oscillations frequencies (b) corresponding to different azimuthal wavenumbers *k* for aspect ratio varying from 0.2 to 1.

Fig. 2. Marginal Reynolds numbers (a) and marginal oscillations frequencies (b) corresponding to different azimuthal wavenumbers *k* for aspect ratio varying from 0.1 to 2.

Fig. 3. Base flow (upper frames) and patterns of the most unstable perturbation (dominant eigenmodes) $\boldsymbol{u} = \{u_r(r,z), u_\theta(r,z), u_z(r,z)\}$ at aspect ratio $\gamma$=0.6 .

Fig. 4. Base flow (upper frames) and patterns of the most unstable perturbation (dominant eigenmodes) $\boldsymbol{u} = \{u_r(r,z), u_\theta(r,z), u_z(r,z)\}$ at aspect ratio $\gamma$=0.25 .

Fig. 5. Base flow (upper frames) and patterns of the most unstable perturbation (dominant eigenmodes) $\boldsymbol{u} = \{u_r(r,z), u_\theta(r,z), u_z(r,z)\}$ at aspect ratio $\gamma$=0.1 .

Table 1. Marginal Reynolds numbers and frequencies at $\gamma = 0.1$ and $15 \leq k \leq 25$

| k  | $Re_m$ | $\omega_m$ |
|----|--------|------------|
| 15 | 23866. | -1.2536    |
| 16 | 23190. | -1.54036   |
| 17 | 22743. | -1.83143   |
| 18 | 22471. | -2.12642   |
| 19 | 22351. | -2.42576   |
| 20 | 22374. | -2.72976   |
| 21 | 22538. | -3.03854   |
| 22 | 22842. | -3.35219   |
| 23 | 23289. | -3.67079   |
| 24 | 23879. | -3.99439   |
| 25 | 24602. | -4.32300   |

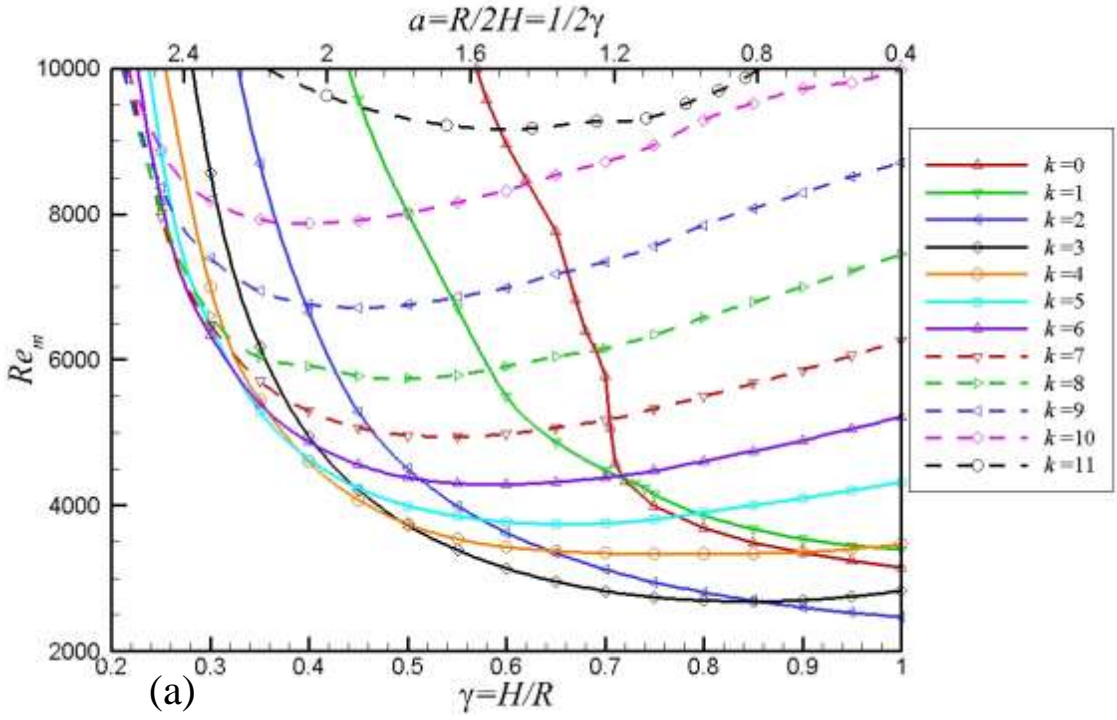
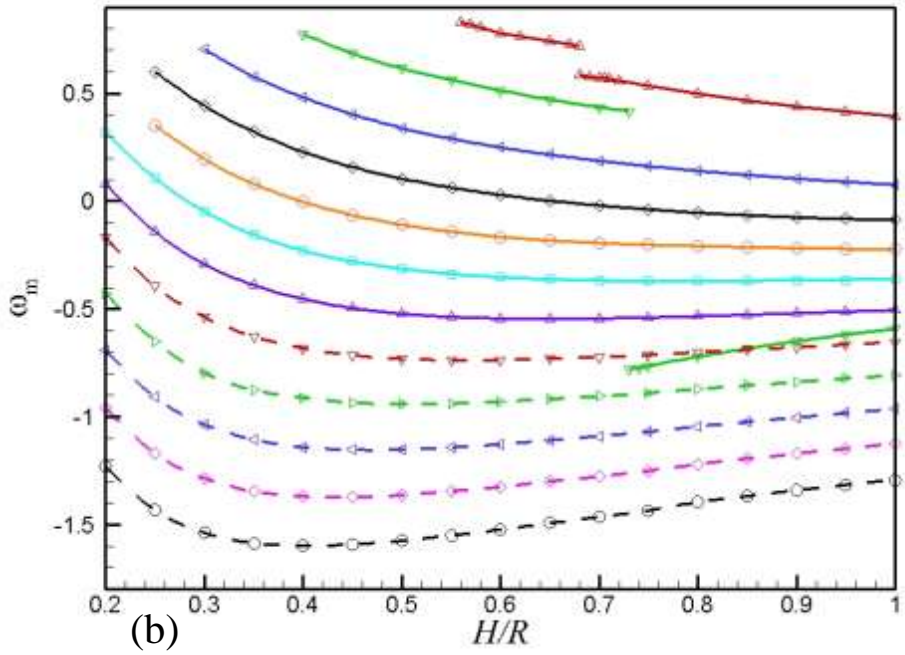

Fig. 1. Marginal Reynolds numbers (a) and marginal oscillations frequencies (b) corresponding to different azimuthal wavenumbers $k$ for aspect ratio varying from 0.2 to 1.

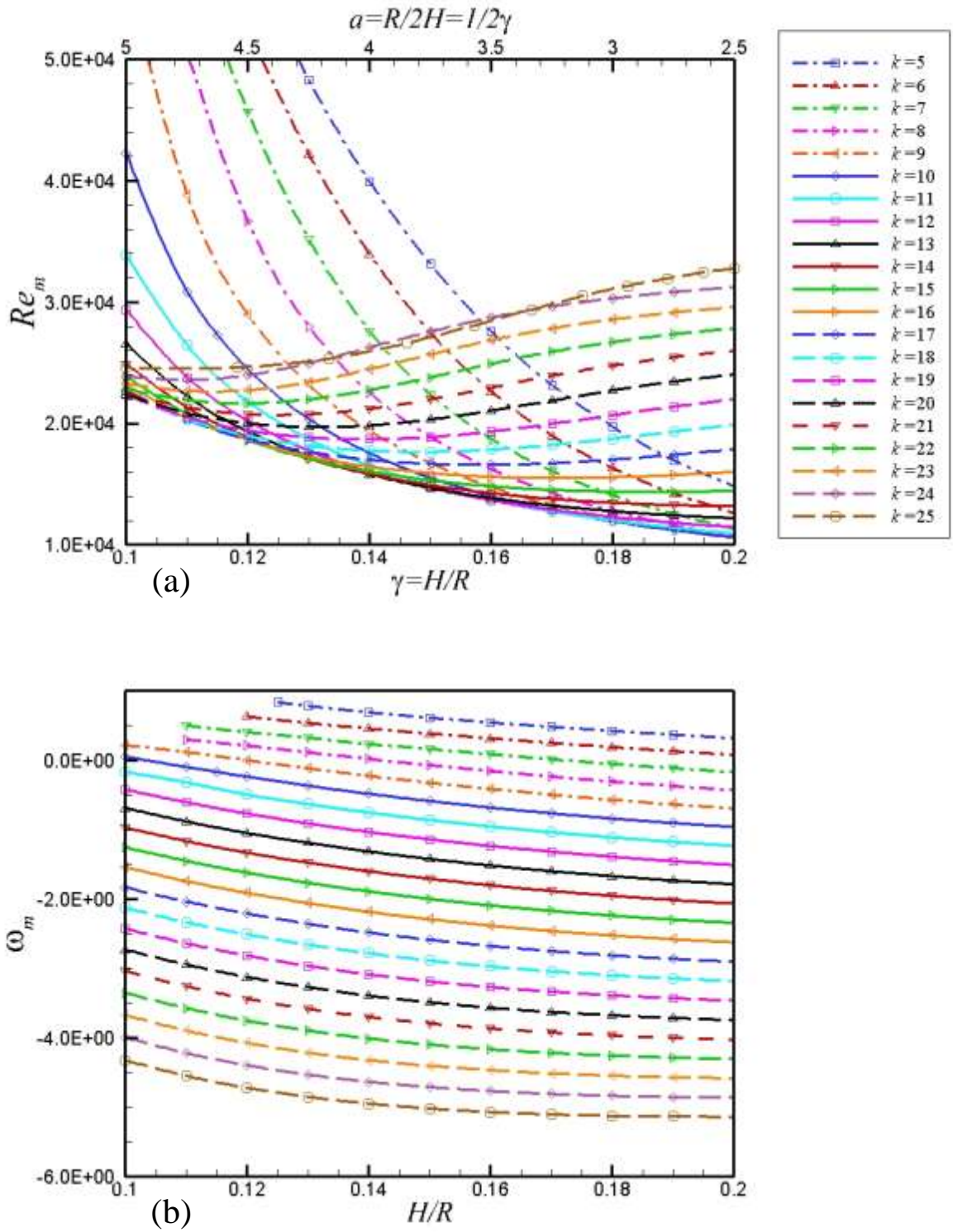

Fig. 2. Marginal Reynolds numbers (a) and marginal oscillations frequencies (b) corresponding to different azimuthal wavenumbers $k$ for aspect ratio varying from 0.1 to 2.

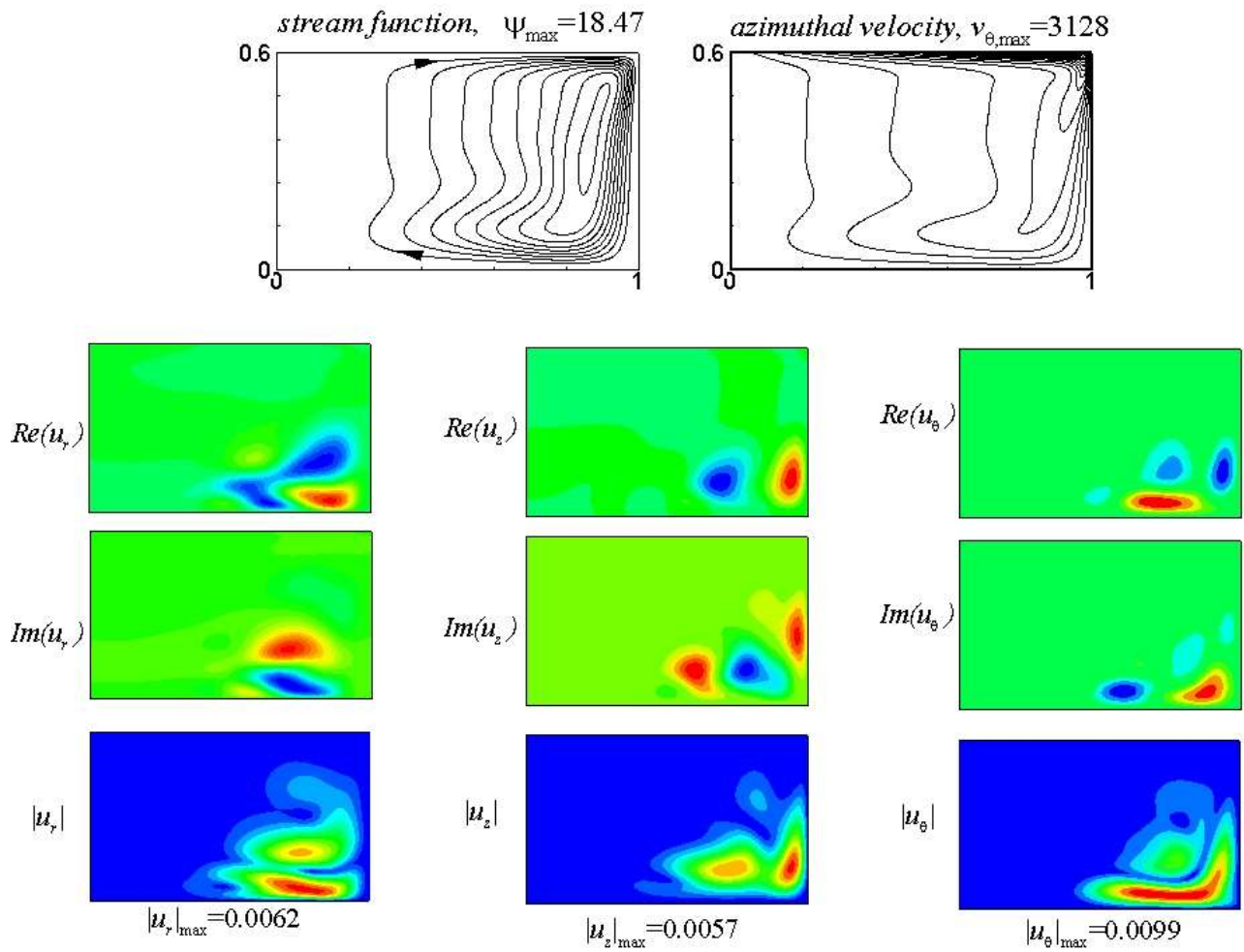

Fig. 3. Base flow (upper frames) and patterns of the most unstable perturbation (dominant eigenmodes) $\boldsymbol{u} = \{u_r(r,z), u_\theta(r,z), u_z(r,z)\}$ at aspect ratio $\gamma=0.6$.

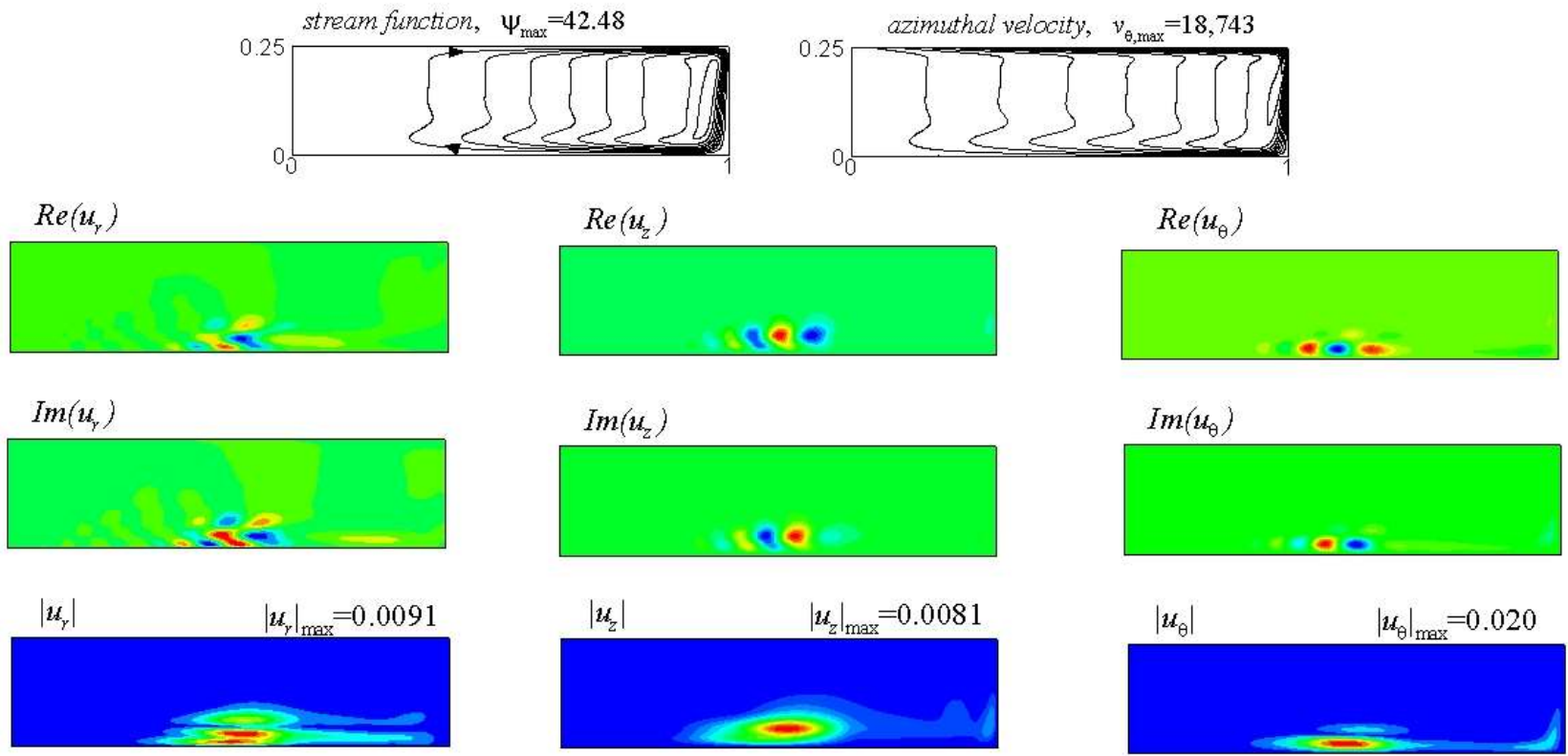

Fig. 4. Base flow (upper frames) and patterns of the most unstable perturbation (dominant eigenmodes) $\boldsymbol{u} = \{u_r(r,z), u_\theta(r,z), u_z(r,z)\}$ at aspect ratio γ=0.25 .

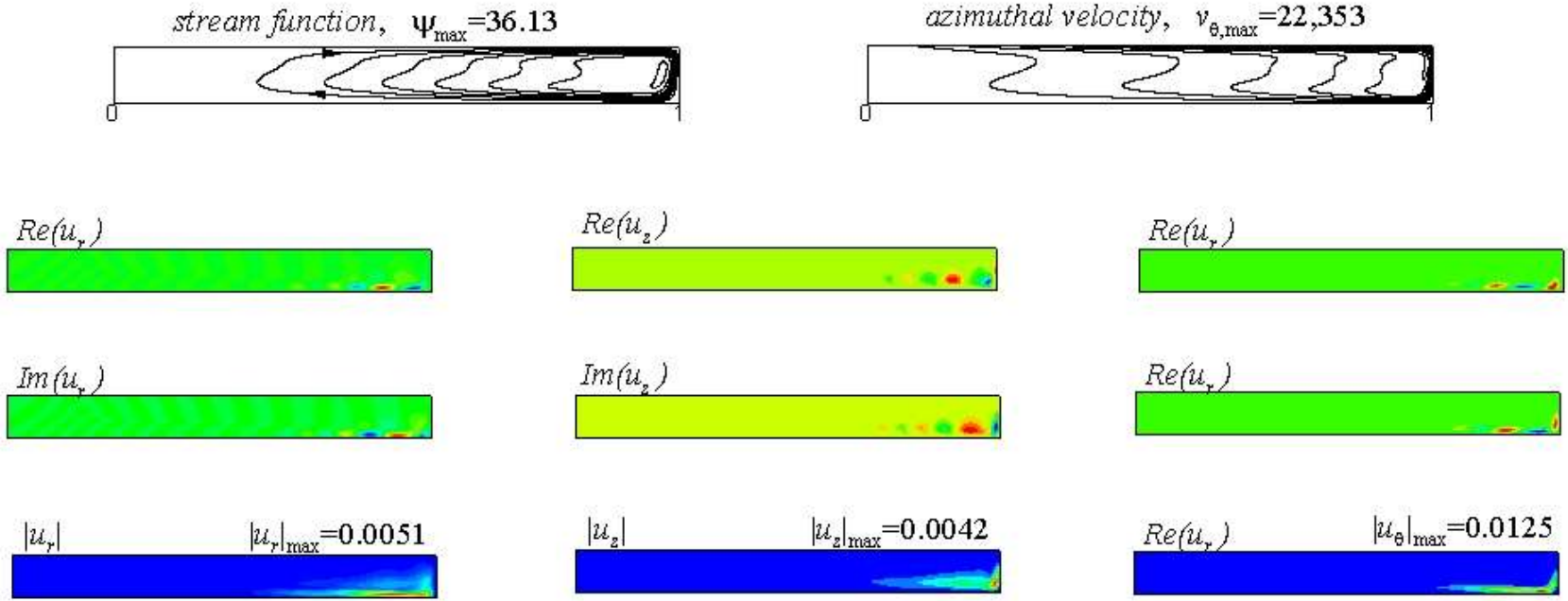

Fig. 5. Base flow (upper frames) and patterns of the most unstable perturbation (dominant eigenmodes) $\boldsymbol{u} = \{u_r(r,z), u_\theta(r,z), u_z(r,z)\}$ at aspect ratio $\gamma=0.1$ .